\documentclass[conference]{IEEEtran}

\ifCLASSINFOpdf 
\else 
\fi

\usepackage{cite}
\usepackage{amsmath,amssymb,amsfonts}
\usepackage{algorithmic}
\usepackage{graphicx}
\usepackage{textcomp}
\usepackage{xcolor}
\usepackage{subcaption}
\usepackage{multicol,multirow}
\usepackage{enumerate,textcomp,array}
\usepackage{placeins}
\usepackage{threeparttable}
\usepackage{dblfloatfix}

\usepackage{booktabs}
\usepackage{balance}

\usepackage{url}

\usepackage[]{fancyhdr} %
\newcommand{\changefont}{\fontsize{7}{7}\selectfont}
\IEEEoverridecommandlockouts
\begin{document}

\title{Droop-e: Exponential Droop as a Function of Power Output for Grid-Forming Inverters with Autonomous Power Sharing}

\author{
R.~W.~Kenyon,~\IEEEmembership{Student~Member,~IEEE}, 
A.~Sajadi,~\IEEEmembership{Senior~Member,~IEEE}, and
B.~M.~Hodge,~\IEEEmembership{Senior~Member,~IEEE}

\thanks{R.~W.~Kenyon and B.~M.~Hodge are with the Department of Electrical, Computer and Energy Engineering at the University of Colorado Boulder, 425 UCB, Boulder, CO 80309, USA, and the Grid Planning and Analysis Center at the National Renewable Energy Laboratory (NREL), 15013 Denver W Pkwy, Golden, CO 80401, USA, Email: \{Richard.Kenyonjr,BriMathias.Hodge\}@colorado.edu, \{Richard.Kenyon,Bri.Mathias.Hodge\}@nrel.gov}

\thanks{A.~Sajadi and B.~M.~Hodge are with the Renewable and Sustainable Energy Institute at the University of Colorado Boulder, 4001 Discovery Drive, Boulder, CO 80303, USA, Email: Amir.Sajadi@colorado.edu}

}


\maketitle
\thispagestyle{fancy}
\pagestyle{fancy}

\begin{abstract}

This paper presents the novel \textit{Droop-e} grid-forming inverter control strategy, which establishes an active power--frequency relationship based on an exponential function of the inverter power dispatch. The advantages of this control strategy include an increased utilization of available headroom, mitigated system frequency dynamics, and a natural limiting behavior, all of which are directly compared to the hitherto standard static droop approach. First, the small signal stability of the \textit{Droop-e} control is assessed on a 3-bus system and found stable across all possible inverter power dispatches. Then, time-domain simulations show improved frequency dynamics at lower power dispatches, and a limiting behavior at higher dispatches. Finally, a novel secondary control scheme is introduced that achieves power sharing following the primary \textit{Droop-e} response to load perturbations, which is shown to be effective in time-domain simulations of the 3- and 9-bus systems; comparative simulations with a static 5\% droop yields unacceptable frequency deviations, highlighting the superiority of the \textit{Droop-e} control.

\end{abstract}


\section{Introduction}
\label{sec:introduction}

As the shares of energy supplied by inverter based resources (IBRs) continues to grow around the world, dynamical challenges associated with the fundamental differences between IBRs and synchronous generators (SGs) become more exacerbated \cite{kenyon_stability_2020,sajadi_integration_2019}. In particular, with the hitherto ubiquitous grid-following (GFL) control approach for parallel connected IBRs, very high instantaneous power penetrations become infeasible due to system dynamics and stability related concerns stemming from a paucity of \textit{grid-forming} (herein understood as devices that establish and generally regulate the local voltage waveform) assets on the power system \cite{ulbig_impact_2014,kenyon_grid-following_2020,sajadi_synchronization_2022}. Thus, attention in both academia and industry has recently shifted towards grid-forming (GFM) IBRs, which regulate the local frequency and voltage magnitude \cite{nerc_grid_2021} independently, as opposed to conventional GFL IBRs that regulate real and reactive power injections as a function of the local voltage and frequency. In this paper, the frequency regulation capability of GFMs is leveraged to devise a novel frequency control method, \textit{Droop-e}, which unlocks the full power potential of GFM devices, enabling the reliable and secure operation of power grids supplied by up to 100\% IBRs, and offering autonomous power sharing amongst all frequency responsive devices.



The presence of a power generation device that directly regulates frequency on a power system is unprecedented; even an SG, the traditional \textit{grid-forming} asset on power systems, has frequency trajectories first and foremost dictated and constrained by the laws of rotational kinematics (i.e., the swing equation). On the contrary, GFM IBRs have substantial control freedom and response agility due to the absence of physical motions and are instead primarily constrained by the limits of power availability and device component ratings (e.g., switches, capacitors, etc). This fundamental contrast between the GFM IBR and the SG presents a great opportunity for a modern approach to generation device enhanced operability, particularly when these GFM devices are paired with storage or curtailed resources, an unavoidable reality during high IBR futures when positive headroom, a fundamental requirement for general power system operation, must be sourced from IBRs. 


A review of the state of the art reveals that a variety of control schemes exist for the GFM approach \cite{unruh_overview_2020}, including static droop \cite{piagi_autonomous_2006}, virtual synchronous machine \cite{beck_virtual_2007}, and virtual oscillator control \cite{johnson_synthesizing_2016}. While these approaches have substantially different dynamical responses, they all center on linear frequency--power relationships. 
Variations on static droop include a feedforward mechanism know as \textit{selfsync}, virtual impedance loop (primarily for reactive power sharing), adaptive droop with a power differential element, and robust droop (applicable to resistive networks) \cite{tayab_review_2017}. Matching control adjusts phasor frequency based on the DC-link capacitor voltage state, which is an indicator for unequal power flow through the device and roughly analogous to the kinetic energy in a synchronous generator rotor \cite{huang_virtual_2017}. Varied droop gains are explored briefly in \cite{lasseter_grid-forming_2020}, with no discussion on the frequency dynamics but rather the steady state impacts and unequal load sharing. The work in \cite{wang_exponential-function-based_2019} explored the concept of an exponential type droop control for distributed energy resources (DERs), but the aim is more accurate reactive power sharing in resistive networks. The exponential relation is inverted as compared to the proposed method herein; whereas the goal in \cite{wang_exponential-function-based_2019} is to not exceed a certain frequency threshold, here the goal is to provide the most power to the network prior to natural limiting. Zhong \cite{zhong_robust_2013} investigates limitations in droop control based on resistive (distributed) impedances and the $P$ -- $E$, $Q$ -- $\omega$ scheme. GFM power overload limiting by rapidly reducing the frequency with a PI controller has been presented in \cite{du_survivability_2019,du_modeling_2019,pattabiraman_comparison_2018}.

This paper introduces a novel GFM control method, \textit{Droop-e}, that leverages the unique capability of GFM inverters to directly regulate frequency as an exponential function of real power output, making it a nonlinear control approach. This is a departure from the convention, the linear droop control which regulates frequency directly proportional to changes in real power output. The \textit{Droop-e} control improvement of system frequency response is multifaceted, including: 1) a larger utilization of available headroom, 2) improved frequency dynamics with exhibiting higher damping, 3) less deviant nadir and a more favorable rate of change of frequency (ROCOF), and 4) an intrinsic, natural power limiting behavior. Additionally, a novel secondary controller is presented that enables power sharing amongst interconnected devices following the primary \textit{Droop-e} response.

\section{Fundamentals of Frequency Response}
\label{sec:motivation}

This section discusses the fundamentals of device frequency response for SGs and GFM IBRs, highlighting the contrasts and motivational basis for the novel \textit{Droop-e} control.


\subsection{Frequency Response of Synchronous Generators}

In conventional power systems, wherein power deficits are compensated by SGs through a governor response, a generation-load imbalance requires a commensurate deviation in frequency as a signal for SG governors to adjust power output. 
The change in power supplied to the network by governor action is a function of frequency, as described in \eqref{eq:SG headroom}:
\begin{equation}
    \label{eq:SG headroom}
    p_{m,G} - p_{m,G,set} = D (\omega_0 - \omega)
\end{equation}
where $p_{m,G}$ is the SG mechanical power, $p_{m,G,set}$ is the exogenous SG mechanical power setpoint, $D$ is the droop gain, which in per unit is 5\% in the United States, $\omega_{0}$ is the radian frequency setpoint, and $\omega$ is the local, and system-wide synchronization radian frequency upon reaching steady state. 
The core governor dynamics of an SG are captured by \eqref{eq:SG governor}, and a basic no-reheat turbine in \eqref{eq:SG turbine}, as presented in \cite{sauer_power_2017}:
\begin{align}
    T_{SV}\frac{dp_{SV}}{dt} &= - p_{SV} + p_{m,G,set} - \frac{1}{D}\left(\frac{\omega_G}{\omega_{s}} - 1\right)\label{eq:SG governor}\\
    T_{CH}\frac{dp_{m,G}}{dt} &= - p_{m,G} + p_{SV}\label{eq:SG turbine}
\end{align}
where $T_{SV}$ is the valve time constant, $p_{SV}$ is the steam chest power command, $D$ is the droop gain, $\omega_G$ is the SG frequency, $\omega_{s}$ is the synchronous frequency, $T_{CH}$ is the turbine steam chest time constant, and $p_{m,G}$ is the mechanical power, equal to the device mechanical torque ($t_{m}$) in per unit. The reciprocal position of $D$ in \eqref{eq:SG governor} shows that for values of $D$ approaching 0\%, the governor dynamics become increasingly faster without bound. The frequency dynamics of the device evolve according to the swing equation \eqref{eq: swing}; the damping component is not shown for illustrative purposes: 
\begin{equation}\label{eq: swing}
    \frac{2H}{\omega_s} \frac{d\omega_G}{dt} = p_{m,G} - p_{e,G}
\end{equation}
where $H$ is the inertia constant of the device and $p_{e,g}$ is the electrical power. Transient load perturbations manifest as deviations in $p_{e,g}$, which cause the frequency to evolve according to \eqref{eq: swing}. Only after the frequency changes will the governor/turbine systems modulate $p_{m,G}$; changes in $p_{m,G}$ due to a perturbation are a function of $\omega_G$, and inversely proportional to $D$. 
To achieve larger $p_{m,G}$ contributions to a relative network perturbation would require smaller $D$ values, which may result in instability due to the increase in rate of change of $p_{SV}$  \eqref{eq:SG governor}, caused by the reciprocal relationship with $D$. Operating at $D=0$ is mathematically infeasible.



\subsection{Frequency Response of Grid-Forming Inverters}

In emerging power systems with more GFM IBRs coming online, the frequency-power dynamic response of power systems might be governed differently. The droop controlled GFM frequency dynamics are shown in \eqref{eq: GFM freq} and \eqref{eq: GFM rocof} \cite{kenyon_open-source_2021}: 
\begin{align}
    \label{eq: GFM freq}\frac{d\delta_{I}}{dt} &=  D \left(p_{m,I,set} - p_{m,I}\right) + \omega_{set}\\
    \label{eq: GFM rocof}\frac{d\omega_I}{dt} &= D\omega_{fil}\left(p_{m,I} - p_{meas,I}\right)
\end{align}
where $\delta_I$ is the inverter electric angle, $D$ is the droop gain, $p_{m,I,set}$ is the exogenous power setpoint, $p_{m,I}$ is the filtered power, $\omega_I$ is the inverter frequency, $\omega_{fil}$ is the power measurement cutoff frequency, and $p_{meas,I}$ is the measured, instantaneous power output. This control approach leverages the natural frequency--droop characteristics of inductive networks to distribute power perturbations amongst devices on the network\cite{mallemaci_comprehensive_2021}. 
Conspicuously absent as a control variable in the frequency dynamics of the GFM is $\omega_I$. Changes in $\omega_I$ and the point of interconnection frequency result in power deviations due to the laws of power flow; in fact, it is appropriate to think that power is extracted from a GFM due to the frequency regulation approach of the device. A change in $\omega_I$ is not required to change the power exported to the network; with respect to frequency regulation, GFM devices are proactive. $D$ is a lever to influence how the local frequency changes, as a function of $p_{m,I}$. Expressed another way, a GFM can deliver larger amounts of power to the network by simply changing the frequency less, which is accomplished with a smaller $D$. As $D\longrightarrow 0$, deviations of $p_{m,I}$ yield decreasing changes in frequency, and the rate of change of frequency (ROCOF), expressed in \eqref{eq: GFM rocof}, eventually reaches zero. The governing equations indicate that operating at $D=0$ is feasible.



\subsection{Comparative Analysis}

\begin{figure}[h]	
	\centering
	\begin{subfigure}[t]{2.1in}
		\centering
		\includegraphics[trim=4 3 20 24,clip,width=1\textwidth]{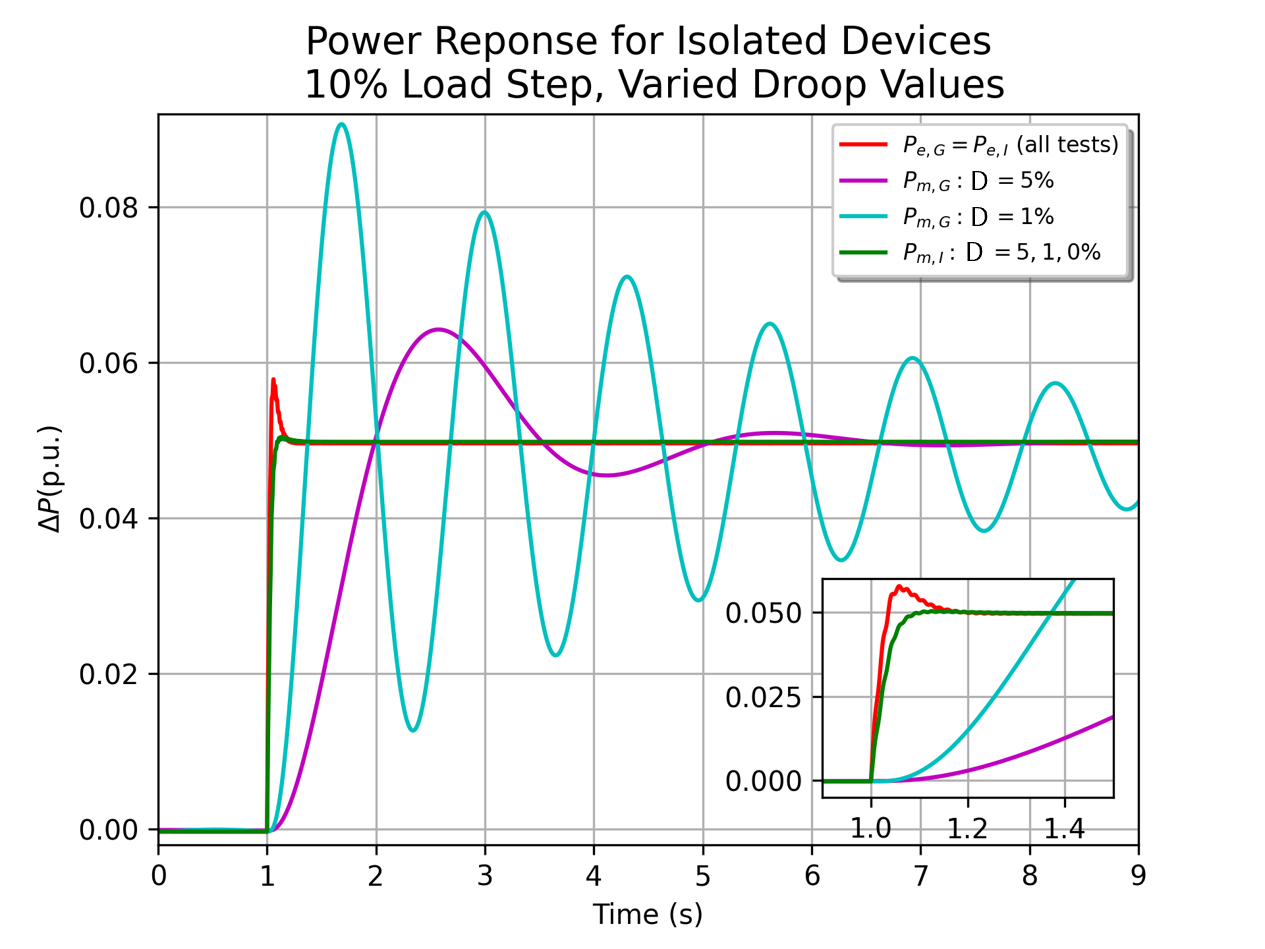}
		\caption{Active Power Response}\label{fig: power varied droop}
	\end{subfigure}
	\quad
	\begin{subfigure}[t]{2.1in}
		\centering
		\includegraphics[trim=2 3 20 24,clip, width=1\textwidth]{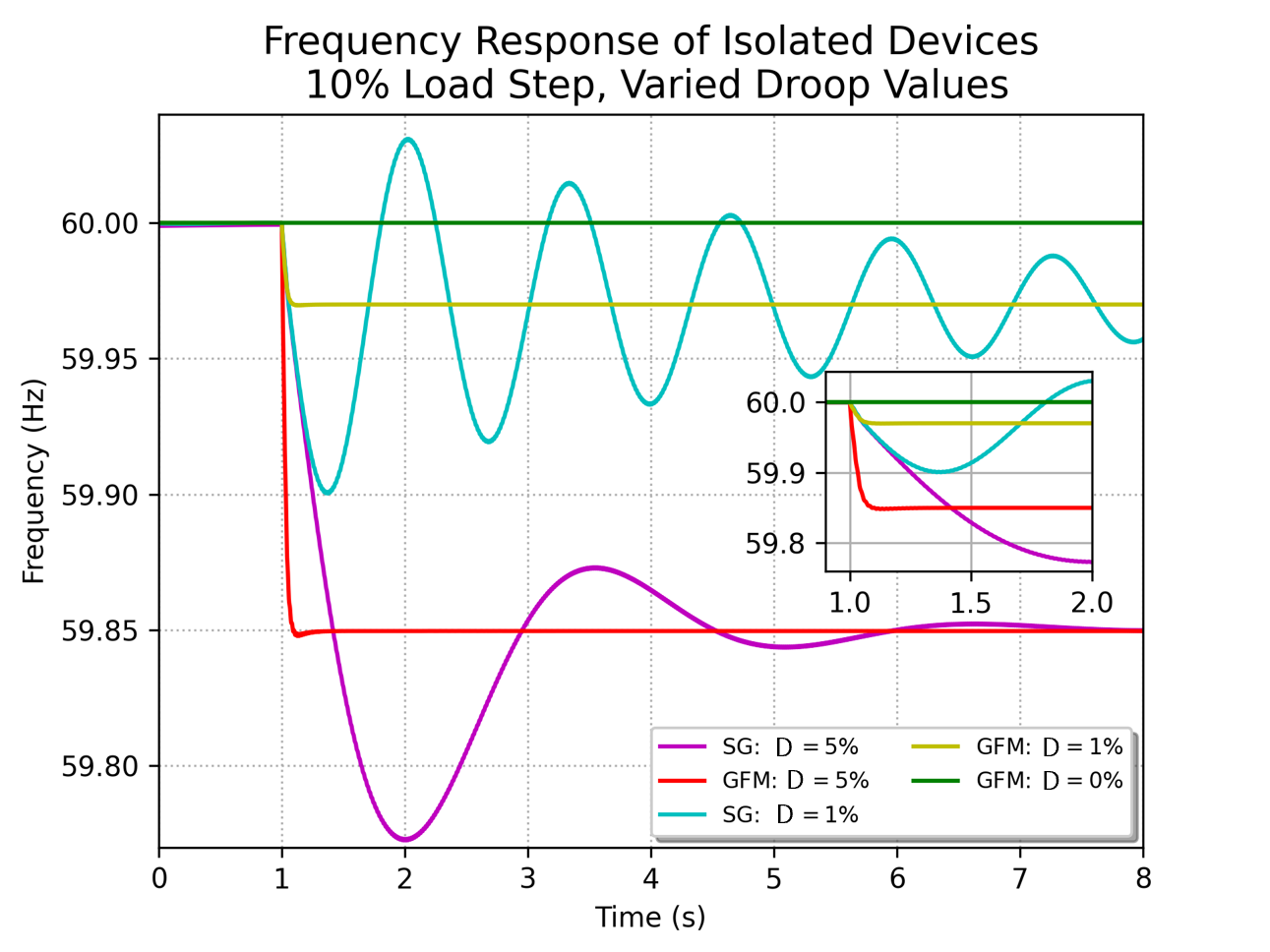}
		\caption{Frequency Response}\label{fig: frequency varied droop}
	\end{subfigure}
	\caption{Power and frequency response of isolated GFM and SG devices to a 10\% load step at a 50\% initial loading. $p_e$ is identical for each device. Varied droop gains are tested.}\label{fig: Power and Freq varied droop}
\end{figure}

To demonstrate the different underlying dynamics for the SG and GFM, simple load step simulations in the power systems computer aided design (PSCAD) software environment were performed. The results are the time-domain response of an SG and a droop-controlled GFM for varied values of $D$. The devices are isolated and connected to a load that is 50\% of the device rating. A 10\% load step perturbation was applied for three different scenarios; $D = 5\%$, $D = 1\%$, $D = 0\%$. The results are shown in Fig. \ref{fig: Power and Freq varied droop}. 

The time-domain plots indicate that the SG response contains an oscillatory mode with a growing frequency for smaller values of $D$. This corroborates the observation of the infeasible operation at $D = 0$ for an SG from \eqref{eq:SG governor}. On the other hand, the GFM response shows no signs of instability for any of the tested values of $D$, indicating that at a local level, $D$ can take any of the simulated values.

\subsection{Observations and Motivation}

The maintenance of a static droop gain across a system is sensible from a load sharing perspective, as all devices will contribute to load deviations proportionally according to the device rating. However, the change in power supplied to the network for frequency responsive devices such as SGs requires a frequency deviation that may be substantial depending on the disturbance; e.g., at a static 5\% droop, a 25\% increase in $p_{m,g}$ requires a 0.75 Hz deviation. For GFM devices at a 5\% static droop gain, the served power incurs a potentially large frequency deviation for what might be a relatively small delivery of available headroom. If a GFM device would change the local frequency less for a given disturbance, achievable with a smaller $D$, a larger amount of power could be extracted (if available), while simultaneously reducing the impact to frequency dynamics. For instance, the nadir for a given load perturbation would be relatively higher, while the rate of change of frequency would be smaller.
The tests from Fig. \ref{fig: Power and Freq varied droop} show that variations in $D$ for an SG are limited, but $D$ can potentially take any value for a GFM inverter. The idea of non-5\% droop gains has been presented \cite{lasseter_grid-forming_2020}, but the shortcomings of a smaller static droop gain include poor transient load sharing, and a high susceptibility to limit violations, particularly at higher dispatches. These shortcomings serve as the motivation for \textit{Droop-e} control, where the simulation supported, hypothesized unbounded limit for assignment of $D$ for the GFM will be leveraged.

\section{The Droop-e Concept}
\label{sec:exponential droop}


The primary idea behind the \textit{Droop-e} concept is making $D$ a function of available headroom, which is accomplished by using $p_{m,I}$ as the independent variable in an exponential (instead of linear) function, as shown in \eqref{eq: exponential droop}:
\begin{equation}\label{eq: exponential droop}
    D = D_e(p_{m,I}) = \omega_b\alpha\left[ e^{\beta(p_{m,I,set})}- e^{\beta(p_{m,I})}\right]
\end{equation}
where $\alpha$ is the proportional scalar, with units of per unit frequency, $\beta$ is the argument scale in per unit power, and $\omega_b$ is the base frequency. This function, $D_e(p_{m,I})$, we call \textit{Droop-e} control. The values of $\alpha$ and $\beta$ have been chosen as 0.002 and 3.0, respectively. 
This paper aims is to serve as a proof of concept and hence, the optimal tuning of these values are not pursued, though future work could focus on any number of constraints such as maximal damping at low dispatches, largest extraction of power prior to under frequency load shedding (UFLS), or acceptable slopes at/near the inverter limits, to name a few. 

\begin{figure}[ht]
    \centering
    \includegraphics[width=1.0\columnwidth,trim={0 0 35 38},clip]{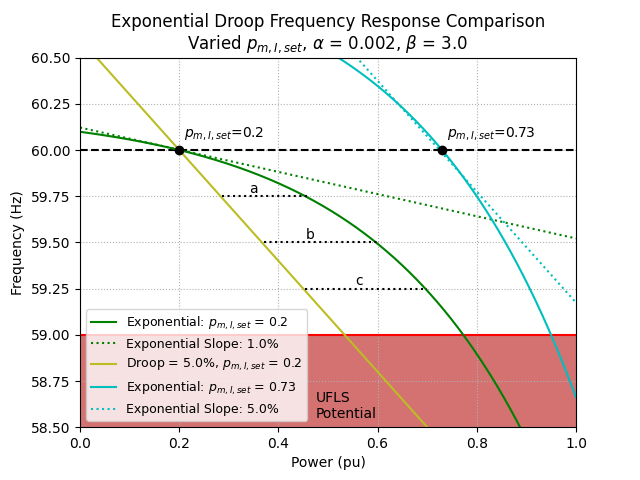}
    \caption{\textit{Droop-e} frequency curves, showing the resultant frequency trajectories for two different dispatches, with tangential droop curves at those dispatches for illumination purposes. A 5\% static droop curve is included for comparison at the $p_{set}=0.2$ dispatch. The region below 59.0 Hz is shaded as an indicator of potential non-linear protection such as UFLS; however, it is noted that some UFLS schemes may trigger at higher frequencies.}
    \label{fig: exponential concept}
\end{figure}

Figure \ref{fig: exponential concept} shows the \textit{Droop-e} power--frequency curves for two different inverter dispatch values, $p_{m,I,set}$, when the steady state frequency is 60 Hz. A network perturbation will cause $p_{m,I}$ to change according to the laws of power flow, because the GFM inverter will initially maintain the local frequency, which incurs changes in angle differentials and a resultant change in power extraction from the inverter. Focusing on $p_{m,I,set}=0.2$, the green trace shows the \textit{Droop-e} frequency trajectory, with a static 5\% droop trajectory (solid yellow trace) also shown at the same dispatch for comparison. The dotted green curve represents an extrapolation of the initial droop value at $p_{m,I,set}=0.2$, which is equal to 1\%. At $p_{m,I,set}=0.73$, due to the values of $\alpha$ and $\beta$ selected for this work, the initial droop value is equal to 5\%. Note the vastly different frequency--power trajectory for the \textit{Droop-e} control at this dispatch, vs. the dispatch $p_{m,I,set}=0.2$. 

The advantage of the proposed control scheme is made obvious when considering the three rays between the \textit{Droop-e} and 5\% static curves from $p_{set}=0.2$, labelled 'a', 'b', and 'c' in Fig. \ref{fig: exponential concept}. The power deviations for each control and resultant frequency deviations are presented in Table \ref{tab: power changes}. Evidently, the \textit{Droop-e} control delivers more power to the network for a given frequency deviation at lower dispatches, which is numerically presented by the $\Delta p_{diff} = (Droop-e) - (Static\phantom{0}5\%)$ values. Thus, \textit{Droop-e} allows the generator to utilize from 18\% to 25\% more of its headroom on a capacity basis than a static linear droop value of 5\%, at $p_{m,I,set}=0.2$, for a 0.75 Hz deviation from the nominal.

\begin{table}[htb]
    \centering
    \caption{Comparison of Power Delivered for \textit{Droop-e} and Static 5\% Control at $p_{m,I,set}=0.2 (pu)$. Corresponds with Fig. \ref{fig: exponential concept}.}
    \label{tab: power changes}
    \begin{tabular}{c|c|c|c|c}
        \multirow{2}{*}{\rotatebox{0}{Ray}} & \multirow{2}{*}{\rotatebox{0}{$\Delta f$ (Hz)}}  & \textit{Droop-e} & Static 5\% & \multirow{2}{*}{\rotatebox{0}{$\Delta p_{diff}(pu)$}} \\
         & & $\Delta p_{m,I}(pu)$ & $\Delta p_{m,I}(pu)$ & \\
        \hline\hline
        a & 0.25 & 0.26 & 0.08 & 0.18 \\
        b & 0.50 & 0.40 & 0.17 & 0.23 \\
        c & 0.75 & 0.50 & 0.25 & 0.25 \\
    \end{tabular}
\end{table}



\subsection{Increased use of Available Headroom}
A primary benefit of the \textit{Droop-e} control is to leverage a larger amount of available headroom for a smaller frequency deviation at relatively lower dispatches, precisely when larger amounts of headroom are available. As a result, the frequency dynamics of the system are suppressed due to the GFM inverter delivering more power to the network with a relatively smaller frequency deviation. While this is helpful to mitigate the dynamics of smaller power systems when load perturbations are on par with the rating of the device, it is also beneficial from a greater headroom delivery potential for larger interconnected systems. 



\subsection{Intrinsic Limiting Capability}
A second benefit to the $Droop-e$ control is the increase in droop slope at higher dispatches. This is advantageous because a GFM inverter cannot export more power than the rating, and a mitigation strategy must be employed. With $Droop-e$ control, the frequency will be lowered at a greater rate at higher dispatches, which will incur larger power extraction from adjacent, frequency responding devices. One type of GFM limiting in the literature is the CERTS limiter \cite{du_survivability_2019}, which employs aggressive PI controllers to rapidly change frequency when violations are met. The benefit of \textit{Droop-e} control over this method is that the device does not enter a non-droop calculated regime with power violations, but instead maintains a droop-type relation with $p_{m,I}$.


\subsection{Lower Rate of Change of Frequency}
A third benefit of the \textit{Droop-e} control comes in the form of reduced ROCOF at lower inverter dispatch levels. The expression of ROCOF in \eqref{eq: GFM rocof} shows a direct proportionality to $D$. With \textit{Droop-e}, this is replaced by $D_e(p_{m,I})$, which is strictly less than $D$ for dispatches below $p_{m,I,set} = 0.73$. Therefore, at these lower dispatches, the ROCOF is less than for a static 5\% droop. This is an important benefit to secure the reliability of power delivery where the grid is equipped with relays that activate on the basis of ROCOF. 




\section{Small Signal Stability Assessment}
\label{sec:SSSA}
The first step in assessing the viability of the \textit{Droop-e} control is a small signal analysis. The small signal stability analysis approach consists of expressing the entire power system including lines, loads, and generators in the differential--algebraic form of \eqref{eq: dynamic variables} and \eqref{eq: algebraic variables}:
\begin{align}
    \frac{dx}{dt} &= f(x,y,u)\label{eq: dynamic variables}\\
    0 &= g(x,y,u)\label{eq: algebraic variables}
\end{align}
where $x$ is a vector of dynamical states, $y$ is a vector of algebraic variables, $u$ is the set of exogenous inputs, $f$ is the set of functions describing the time evolution of the dynamical states, $x$, and $g$ is the set of functions relating the network algebraic variables. \eqref{eq: dynamic variables} and \eqref{eq: algebraic variables} can be linearized in the following form:
\begin{equation}\label{eq: linearized system}
    \Delta \dot{x} = A_{sys}\Delta x + B\Delta u
\end{equation}
where $A_{sys}$ represents the aggregation of all algebraic equations within the dynamical expressions, and $B$ is the matrix of exogenous control parameters. The eigenvalues $\lambda_i$ of $A_{sys}$ are generally complex in the form of $\lambda_i = \alpha_i + j\omega_i$, where $\alpha_i$ and $\omega_i$ are the real and imaginary parts, respectively, of the $i$th eigenvalue. Positive values of $\alpha_i$ indicate fundamental instabilities, while the damping ($\zeta$) of the eigenvalues is calculated as \eqref{eq: damping}:
\begin{equation}\label{eq: damping}
    \zeta_i = \frac{-\alpha_i}{\sqrt{\alpha_i^2 + \omega_i^2}}
\end{equation}

\begin{figure}[h]
    \centering
    \includegraphics[width=0.8\columnwidth,trim={0 0 0 0},clip]{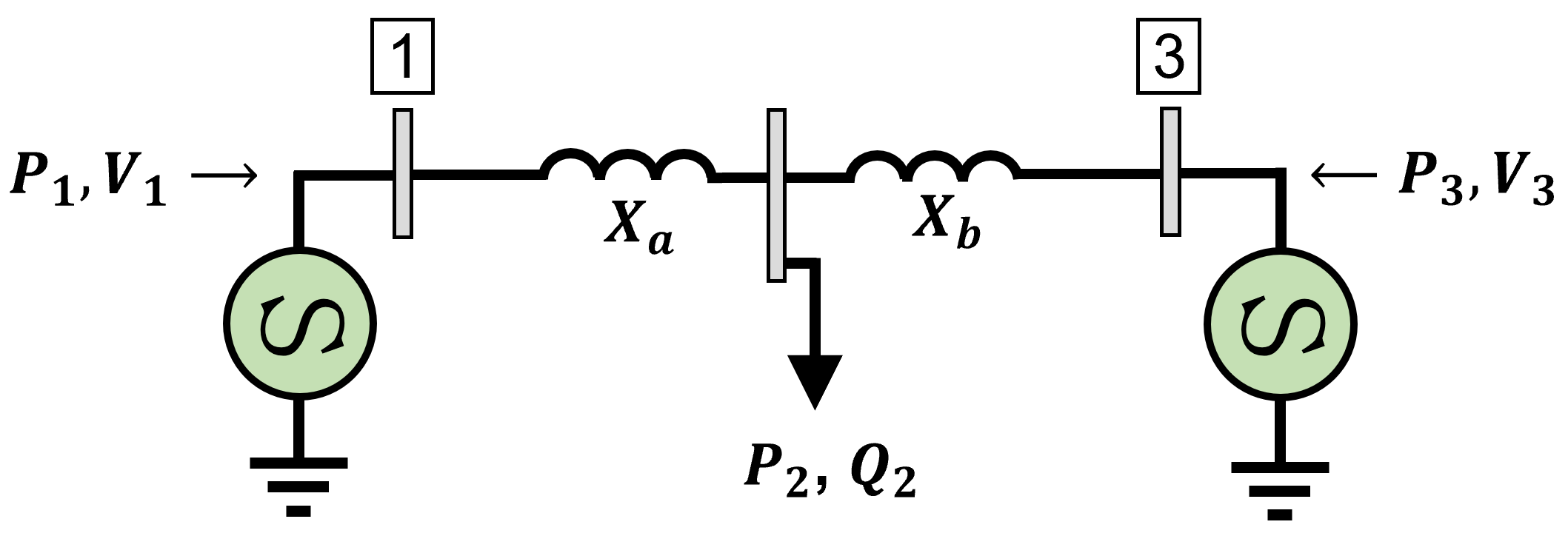}
    \caption{A simple 3-bus system. A synchronous generator is located at bus 1, while a \textit{Droop-e} grid-forming inverter is at bus 3.}
    \label{fig: three bus model}
\end{figure}

Consider the simple 3-bus network of Fig. \ref{fig: three bus model}. This system is used to demonstrate the device level stability via a small signal stability analysis. An SG is located at bus 1 and a \textit{Droop-e} GFM IBR (assumed a battery energy storage system with no energy availability constraints) is at bus 3. The impedances $X_a$ and $X_b$ connect the three buses. The network base is 100 MVA, 18 kV, which applies to all per unit values except for the GFM, which is rated at 50 MVA. The GFM was purposefully chosen at a lower rating as compared to the SG, to show the stabilizing benefit of the \textit{Droop-e} control even at relatively lower ratings. The load at bus 2 is constant power, with a 0.95 leading power factor. The network details are provided in Table \ref{tab: system parameters}.
 

\subsection{Synchronous Generator Model}
\label{sec: SG}

\begin{figure}[ht]
    \centering
    \includegraphics[width=1.0\columnwidth,trim={0 0 0 0},clip]{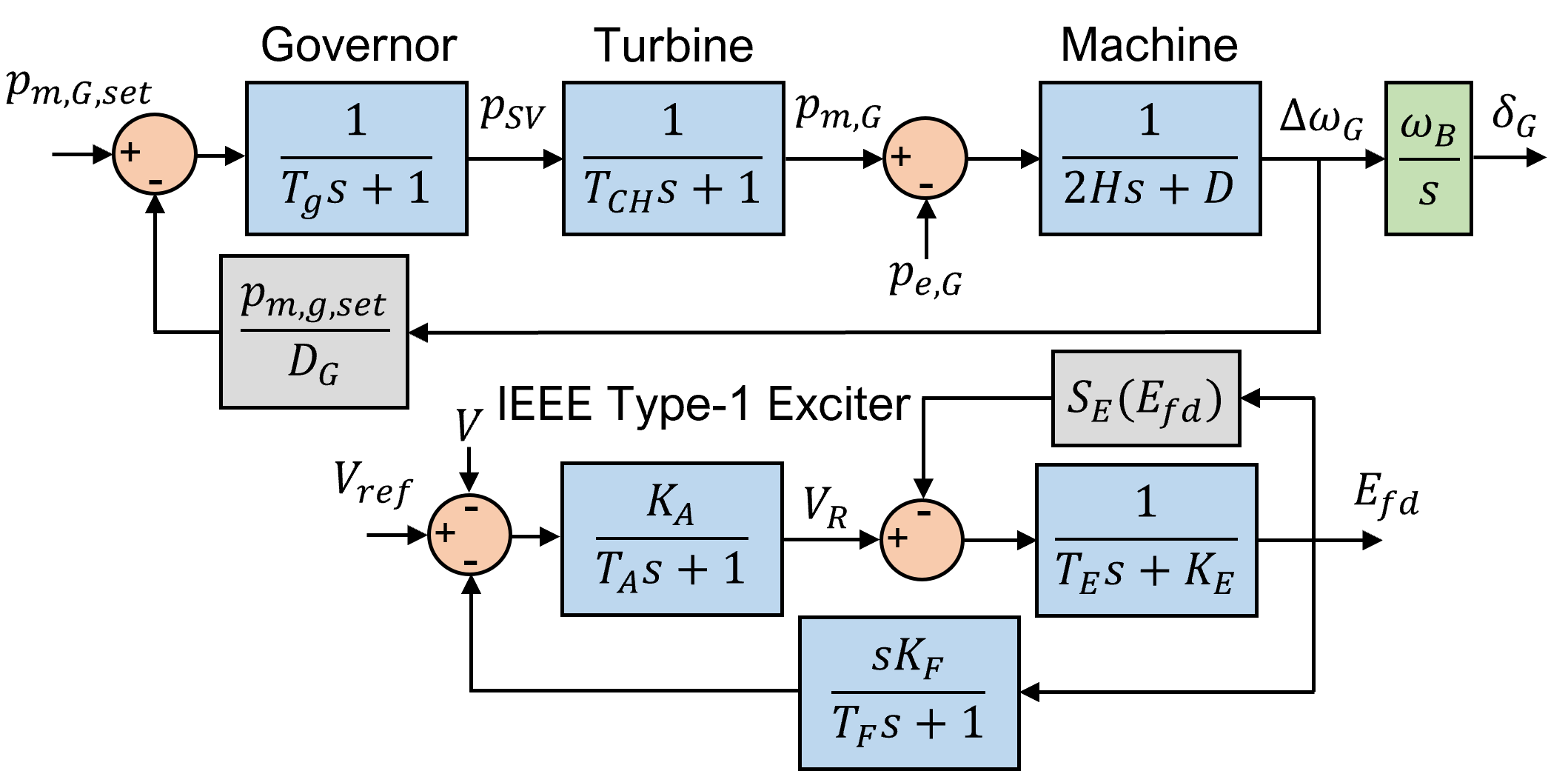}
    \caption{The synchronous generator model with governor, turbine, machine, and exciter dynamic sub-systems.}
    \label{fig: SG Block}
\end{figure}

The SG model used in these studies is constructed from the base model used in \cite{sauer_power_2017}, implemented as shown in the block diagram of Fig. \ref{fig: SG Block}. The governor model \eqref{eq:SG governor} is a first order system acting on the difference between $p_{m,G,set}$ and the droop relation to frequency deviations, $\Delta\omega_G$. The turbine model is a simple steam chest with no reheat process \eqref{eq:SG turbine}. The standard swing equation \eqref{eq: swing} machine dynamics are included. The exciter is based on the IEEE Type-1 model. The saturation function is an exponential of the form: $S_E(E_{fd}) = \gamma e^{\epsilon E_{fd}}$. Flux decay is modelled but not shown in Fig. \ref{fig: SG Block}. The result is a 9-th order model, with the states provided in \eqref{eq: SG states}. The parameter values are those from machine 3 in the 9 bus model developed in \cite{sauer_power_2017}, with governor and turbine parameters from \cite{pattabiraman_comparison_2018}. The standard voltage behind reactance model is used to connect the SG to the network (see \cite{sauer_power_2017}). The SG parameters are provided in Table \ref{tab: system parameters}.

\begin{figure*}[ht]	 
	\centering
	\begin{subfigure}[t]{2.21in}
		\centering
		\captionsetup{justification=centering}
		\includegraphics[trim=3 4 110 40, clip,width=1\textwidth]{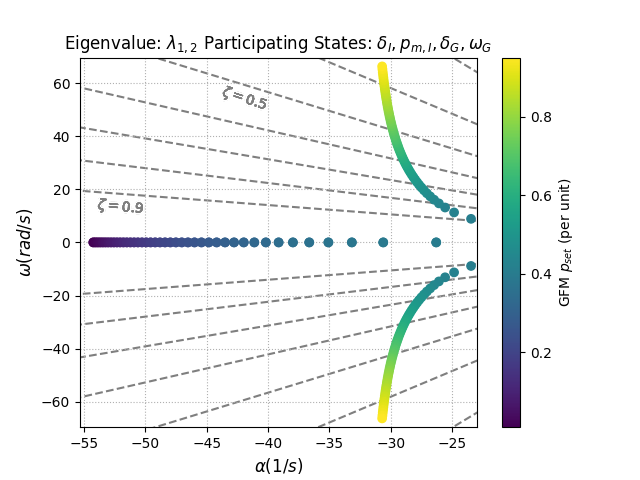}
	    \caption{Eigenvalue: $\lambda_{1,2}$\\ States: $\delta_I$, $p_{m,I}$, $\delta_G$, and $\omega_G$}\label{fig: eigen 1}
    \end{subfigure}
	\hfill
	\begin{subfigure}[t]{2.21in}
		\centering
		\captionsetup{justification=centering}
		\includegraphics[trim=3 4 110 40,clip, width=1\textwidth]{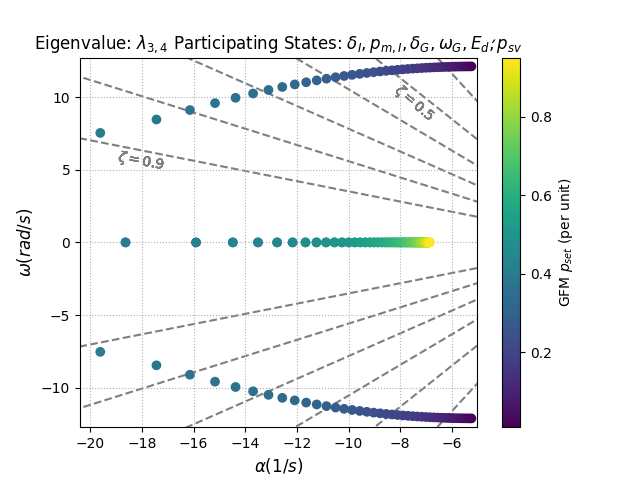}
		\caption{Eigenvalue: $\lambda_{3,4}$\\ States: $\delta_I$, $p_{m,I}$, $\delta_G$, $\omega_G$, $E'_d$, and $p_{SV}$.}\label{fig: eigen 2}
	\end{subfigure}
	\hfill
	\begin{subfigure}[t]{2.62in}
	    \centering
	    \captionsetup{justification=centering}
	    \includegraphics[trim=3 4 40 40,clip, width=1\textwidth]{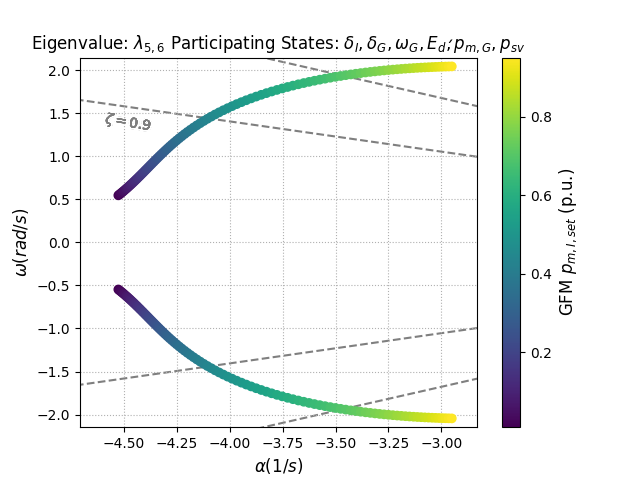}
	    \caption{Eigenvalue: $\lambda_{5,6}$\\ States: $\delta_I$, $\delta_G$, $\omega_G$, $E'_d$, $p_{m,G}$, and $p_{SV}$}\label{fig: eigen 3}
	\end{subfigure}
	\caption{Eigenvalue trajectories of the 3 bus system for those with participation from grid-forming inverter states. Note the varied x and y axis scales.}\label{fig: eigenvalues}
\end{figure*}
\begin{equation}\label{eq: SG states}
    x_{SG} = [\delta_{G}, \omega_{G}, E'_q, E'_d, E_{fd}, V_R, R_f, p_{m,G}, p_{SV}]
\end{equation}


\subsection{Grid-Forming Inverter Model}
\label{sec: gfm droop-e}

\begin{figure}[h]
\centering
    \includegraphics[width=0.8\columnwidth,trim={0 0 0 0},clip]{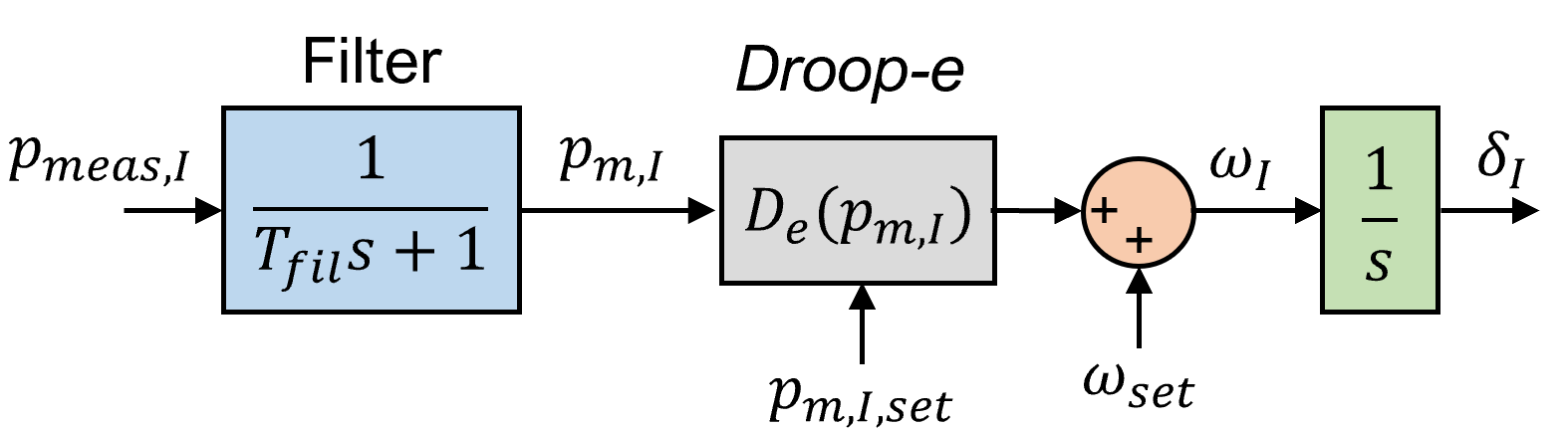}
    \caption{Frequency control with the \textit{Droop-e} method. The instantaneous, measured output power ($p_{meas,I}$) is filtered prior to being passed to the \textit{Droop-e} controller. Integration of the frequency ($\omega_I$) yields the local angle for the GFM ($\delta_I$).}
    \label{fig: exponential droop block}
\end{figure}

The frequency control for the GFM model is shown in Fig. \ref{fig: exponential droop block}. The instantaneous measured power $p_{meas,I}$ is passed through a first order filter with time constant $T_{fil}$. The resultant $p_{m,I}$ value is provided to the \textit{Droop-e} block along with $p_{m,I,set}$ to determine the output frequency, $\omega_I$; a factor of $2\pi$ is not explicitly shown.

\begin{figure}[h]
    \centering
    \includegraphics[width=0.65\columnwidth,trim={0 0 0 0},clip]{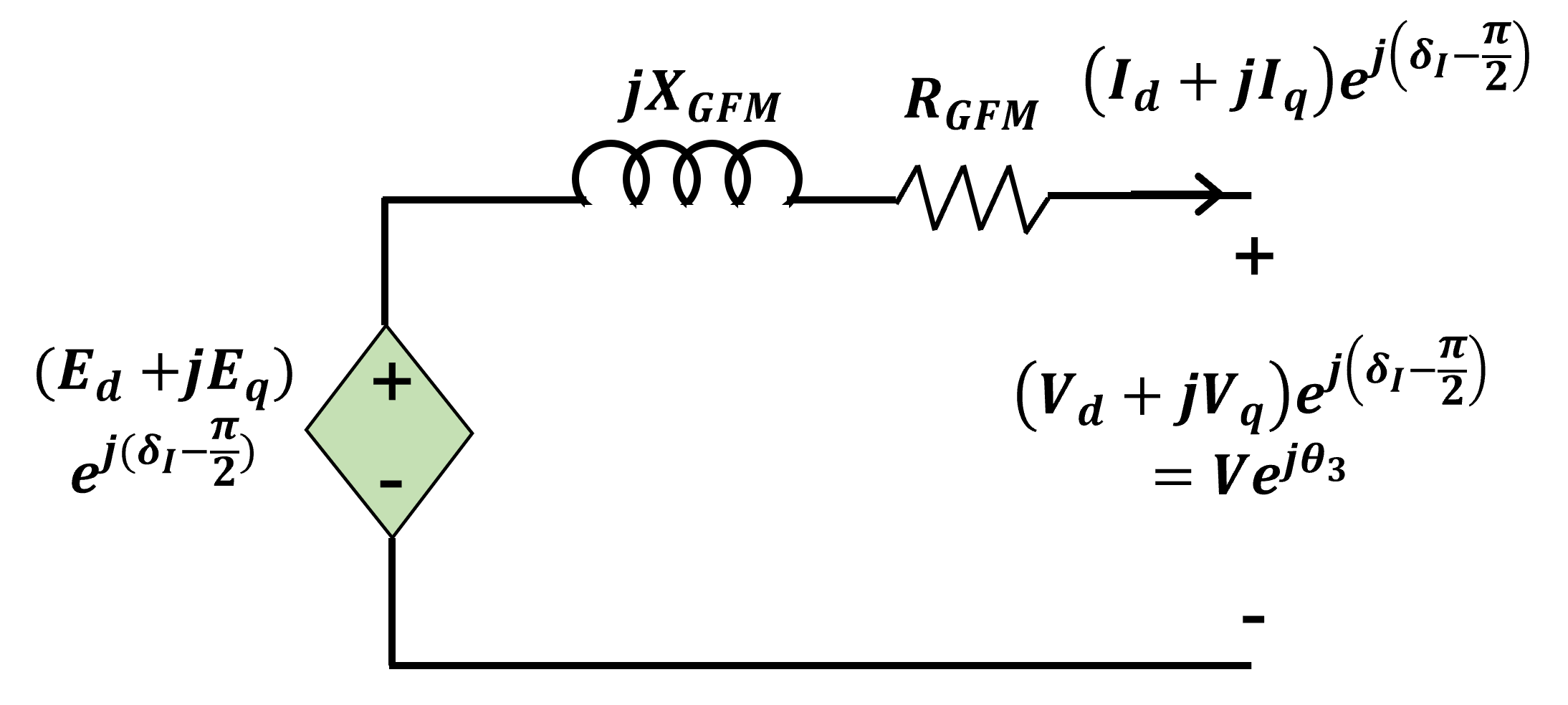}
    \caption{The voltage behind impedance model adopted to represent the grid-forming inverter in the small signal stability analysis.}
    \label{fig:GFM Stator}
\end{figure}

A voltage behind impedance model is used, as shown in Fig. \ref{fig:GFM Stator}, wherein the standard LCL filter coupling inductance is the impedance. A GFM inverter regulates the voltage across the LCL capacitor, which is the voltage provided to the source in Fig. \ref{fig:GFM Stator}. A constant voltage is assumed, which absolves the voltage and current proportional--integral controllers and the filter capacitor and inductor dynamical states \cite{kenyon_open-source_2021}. As the interest here is primarily on the relatively slower frequency dynamics, the constant voltage is practical and similar reductions have been exercised in other analyses \cite{pattabiraman_comparison_2018}. The governing equations of the GFM inverter with \textit{Droop-e} control, as installed at bus 3 in the network of Fig. \ref{fig: three bus model}, are \eqref{eq:gfm angle} and \eqref{eq:gfm pm}:
\begin{align}
    \frac{d\delta_I}{dt} &= \omega_b\alpha\left[ e^{\beta(p_{m,I,set})} -  e^{\beta(p_{m,I})}\right] + \omega_{set}\label{eq:gfm angle}\\
    \frac{d p_{m,I}}{dt} &= \frac{-p_{m,I}}{T_{fil}} + \frac{V_3sin(\delta_I-\theta_3)I_{I,d} + V_{3}cos(\delta_I-\theta_3)I_{I,q}}{T_{fil}}\label{eq:gfm pm}
\end{align}
where $V_3$ is the RMS voltage at bus 3, $\theta_3$ is the angle of bus 3, $\delta_I$ is the internal angle of the GFM, and $I_{I,d}$ and $I_{I,q}$ are the internal $d$ and $q$ axis currents. The internal values are brought into the global reference frame with the $e^{j\left(\delta_I - \frac{\pi}{2}\right)}$ expression. The internal voltages, $E_d$ and $E_q$, are taken as constants. This constant voltage assumption reduces the prototypical 13th order GFM model \cite{kenyon_open-source_2021} to a 2nd order model with the states of \eqref{eq: gfm state}, because the current and voltage controllers, the filter inductor and capacitors, and the reactive power equations, are ignored. The relevant parameters are provided in Table \ref{tab: system parameters}.

\begin{equation}\label{eq: gfm state}
    x_{GFM}= [\delta_{I}, p_{m,I}]
\end{equation}

\begin{table}[h]
    \renewcommand{\arraystretch}{1.0}
    \setlength{\tabcolsep}{.3em}
    \centering
    \caption{System Parameters}
    \begin{tabular}{c|c||c|c||c|c}
    \textbf{Parameter} & \textbf{Value} & \textbf{Par...} & \textbf{Value} & \textbf{Par...} & \textbf{Value}\\\hline\hline
    $H(secs)$ & 3.01 & $X_d(pu)$ & 1.3125 & $X'_d(pu)$ & 0.1813\\\hline
    $X_q$ & 1.2578 & $X'_q(pu)$ & 0.25 & $T'_{do}(sec)$ & 5.89\\\hline
    $T'_{qo}(sec)$ & 0.6 & $K_A$ & 20 & $T_A(sec)$ & 0.2\\\hline
    $K_E$ & 1.0 & $T_E(sec)$ & 0.314 & $K_F$ & 0.063\\\hline
    $T_F(sec)$ & 0.35 & $S_E-\gamma$ & 0.0039 & $S_E-\epsilon$ & 1.555\\\hline
    $D_G(\%)$ & 5 & $\omega_B(rad/s)$ & 377 & $X_a(pu)$ & 0.05\\\hline
    $X_a(pu)$ & 0.05 & $D_I-\alpha$ & 0.002 & $D_I-\beta$ & 3.0\\\hline
    $X_{GFM}(pu)$ & 0.15 & $R_{GFM}(pu)$ & 0.005 & $T_{fil}(sec)$ & 0.0167\\\hline
    $P_2(pu)$ & 0.75 & $Q_2(pu)$ & 0.25 & $V_1(pu)$ & 1.02\\\hline
    $V_3(pu)$ & 1.02 & $S_{G}(MVA)$ & 100 & $S_{I}(MVA)$ & 50\\
    \end{tabular}
    \label{tab: system parameters}
\end{table}

\subsection{Results and Discussion}
The eigenvalues of the 3-bus system of Fig. \ref{fig: three bus model} were calculated for a range of power flows that span the per unit dispatch of the GFM inverter, from 0.01 to 0.99 to assess the stability of the full range of power dispatches for the \textit{Droop-e} controller. Three complex eigenvalue pairs were identified via participation factor analysis as involving the GFM states of \eqref{eq: gfm state}; $\lambda_{1,2}$, $\lambda_{3,4}$, and $\lambda_{5,6}$. The eigenvalue pair $\lambda_{1,2}$ involves the states $\delta_I$, $p_{m,I}$, $\delta_G$, and $\omega_G$. The level of participation varies as the dispatch, but all four states are present through the range of investigated dispatches. Figure \ref{fig: eigen 1} shows that the eigenvalue trajectory is pure real and negative, for the range $p_{set} = [0.01 - 0.4]$, and at $p_{m,I,set} =0.4$ a bifurcation occurs and the modes become oscillatory. The damping, as interpreted by the damping traces of $\zeta$ included in the charts, decreases monotonically up to the full dispatch point, $p_{set} = 0.99$, but remains at a reasonable level.

Figure \ref{fig: eigen 2} depicts the loci of migration of the eigenvalue pair $\lambda_{3,4}$ in the complex plane, which includes participation from the states $\delta_I$, $p_{m,I}$, $\delta_G$, $\omega_G$, $E'_d$, and $p_{SV}$. This eigenvalue pair has an oscillatory element at low $p_{m,I,set}$ values that becomes increasingly damped as the dispatch is increased. The mode becomes pure real at the same bifurcation point as $\lambda_{1,2}$, indicating a likely coupling between the two. The trajectory of this pair shows that the mode is always damped, due to the left half plane location. The eigenvalue pair, $\lambda_{5,6}$, shown in Fig. \ref{fig: eigen 3}, depicts an oscillatory mode at all dispatches, with a decreasing damping as the dispatch of the GFM inverter is increased. This mode is always well damped, with $\zeta < 0.7$ for all dispatches. This mode, which is in the range of $0.15\to0.63 Hz$, is a low frequency oscillatory mode. The trajectory of this pair corroborates the frequency damping benefit to the system of the \textit{Droop-e} control, particularly at low dispatches. 

The results of the small signal stability analysis here suggest that all modes of the 3 bus system with the \textit{Droop-e} GFM control, including those not shown but only involving SG states, have a negative real part and positive damping due to $\alpha_i < 0$ for all $p_{set}$ values, and hence form a stable system.

\begin{figure*}[h]	
	\centering
	\begin{subfigure}[t]{2.35in}
		\centering
		\includegraphics[trim=3 6 43 41,clip,width=1\textwidth]{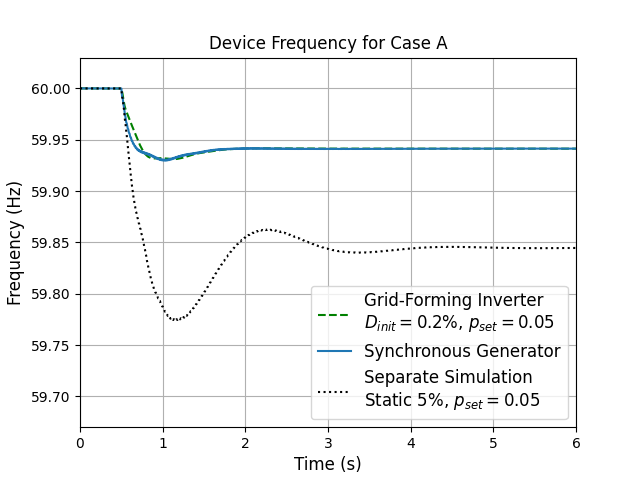}
		\caption{Case A}\label{fig: freq a}		
	\end{subfigure}
    \hfill
	\begin{subfigure}[t]{2.35in}
		\centering
		\includegraphics[trim=3 6 43 41,clip, width=1\textwidth]{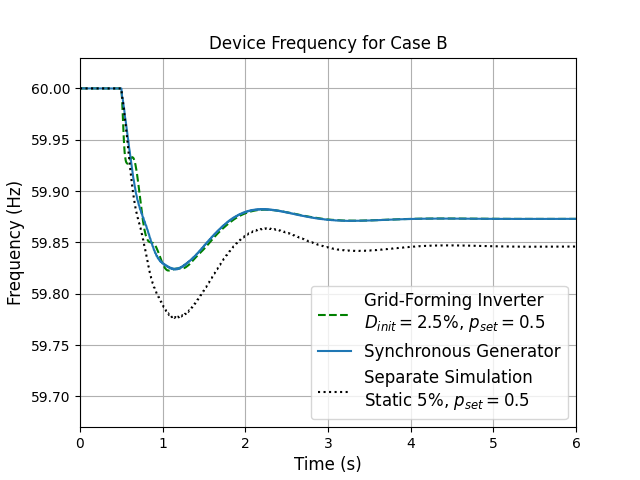}
		\caption{Case B}\label{fig: freq b}
	\end{subfigure}
	\hfill
	\begin{subfigure}[t]{2.35in}
	    \centering
	    \includegraphics[trim=3 6 43 41,clip, width=1\textwidth]{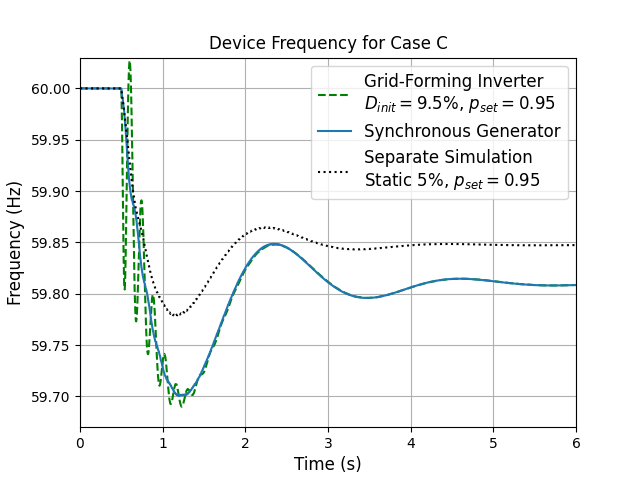}
	    \caption{Case C}\label{fig: freq c}
	\end{subfigure}
	\caption{Device frequency response for the three dispatch cases: A, B, and C.}\label{fig: Frequency for cases}
\end{figure*}

\begin{figure*}[h]	
	\centering
	\begin{subfigure}[t]{2.35in}
		\centering
		\includegraphics[trim=3 6 43 41,clip,width=1\textwidth]{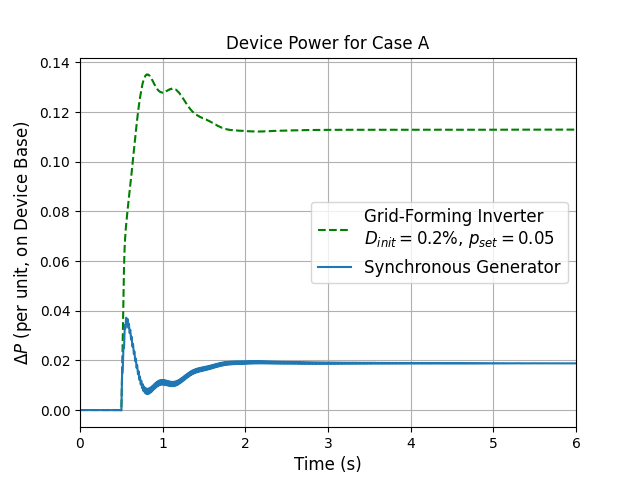}
		\caption{Case A}\label{fig: power a}		
	\end{subfigure}
	\hfill
	\begin{subfigure}[t]{2.35in}
		\centering
		\includegraphics[trim=3 6 43 41,clip, width=1\textwidth]{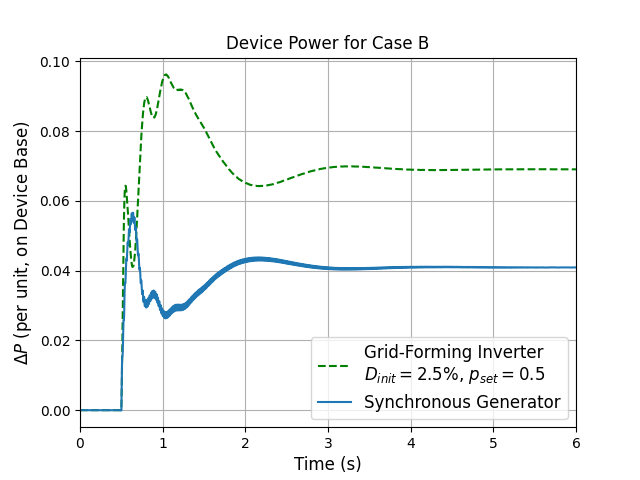}
		\caption{Case B}\label{fig: power b}
	\end{subfigure}
	\hfill
	\begin{subfigure}[t]{2.35in}
	    \centering
	    \includegraphics[trim=1 6 43 41,clip, width=1\textwidth]{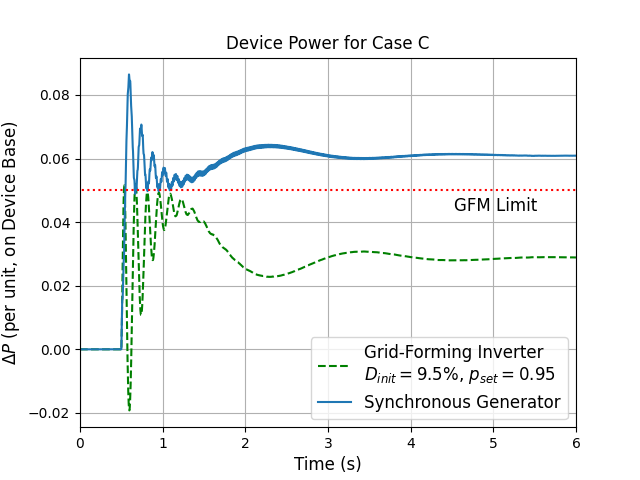}
	    \caption{Case C}\label{fig: power c}
	\end{subfigure}
	\caption{Device real power response for the three dispatch cases; A, B, and C.}\label{fig: Power for cases}
\end{figure*}


\section{Time-Domain Simulations: 3-Bus System}
\label{sec:simulation results}

The dynamic simulations were conducted in the PSCAD software package was used to run dynamic simulations for varied dispatch points to display the full expression of characteristics. The GFM model is 13th order, with internal current and voltage controllers and an output LCL filter. It is executing a reactive power-droop relationship ($V-Q$). A full description of the model can be found in \cite{kenyon_open-source_2021}, while the parameters are those from Tables \ref{tab: system parameters} and \ref{tab: PSCAD system parameters}. The SG model was constructed with internal PSCAD models and established with the parameters from Table \ref{tab: system parameters}. The dynamic models and the 3-bus network used is available open-source at \cite{kenyon_pypscad_2020}.

\begin{table}[h]
    \renewcommand{\arraystretch}{1.2}
    \setlength{\tabcolsep}{.4em}
    \centering
    \caption{Additional Dynamic System Parameters}
    \begin{tabular}{c|c||c|c||c|c}
    \textbf{Parameter} & \textbf{Value} & \textbf{Par...} & \textbf{Value} & \textbf{Par...} & \textbf{Value}\\\hline\hline
    $L_f(pu)$ & 0.15 & $R_f(pu)$ & 0.005 & $C_f(pu)$ & 2.5\\\hline
    $R_{cap}(pu)$ & 0.005 & $k_C^i$ & 1.19 & $k_C^p$ & 0.73\\\hline
    $G_C$ & 1.0 & $k_V^i$ & 1.16 & $k_V^p$ & 0.52 \\\hline
    $G_V$ & 1.0 & frito & n/a & burrito & n/a\\
    \end{tabular}
    \label{tab: PSCAD system parameters}
\end{table}

Three 10\% load step (7.5 MW, 2.5 Mvar) simulations were carried out, where the dispatch of the GFM is varied in order to showcase the improved response at lower dispatches. The initial power flow for each case, 'A', 'B', and 'C', is presented in Table \ref{tab: cases}. Because the SG is rated at 100 MVA, the full range of dispatch with the 50 MVA GFM inverter is possible with the 75 MW load. The simulations are executed by establishing the two devices as ideal sources with no dynamics enabled at the power flow determined voltage and angle. The dynamics are subsequently released in a manner conducive to achieving the desired steady state. Two metrics are used to quantify frequency response; the nadir, which is the lowest frequency value post disturbance, and the peak ROCOF, which is calculated with a sliding window of $T_w = 0.1s$;  $\label{eq:ROCOF} max|\dot{f_G}(t)| = \frac{\omega_G(t + T_{w}) - \omega_G(t)}{2\pi T_{w}}$. The frequency statistics are calculated from the SG rotational frequency alone.

\begin{table}[tb]
    \renewcommand{\arraystretch}{1.2}
    \setlength{\tabcolsep}{.3em}
    \centering
    \caption{Dispatches for each case. Synchronous Generator is 100 MVA; Grid-Forming Inverter is 50 MVA}
    \label{tab: cases}
    \begin{tabular}{c||c|c||c|c}
         \multirow{2}{*}{\rotatebox{0}{\textbf{Case}}} & \multicolumn{2}{c||}{\textbf{Synchronous Generator}} & \multicolumn{2}{c}{\textbf{Grid-Forming Inverter}}\\
         & \phantom{00}\textbf{P (pu)}\phantom{00} & \textbf{Q (pu)} & \phantom{00}\textbf{P (pu)}\phantom{00} & \textbf{Q (pu)} \\
        \hline\hline
        A & 0.73 & 0.25 & 0.05 & 0.13 \\
        B & 0.50 & 0.23 & 0.50 & 0.08 \\
        C & 0.27 & 0.25 & 0.95 & 0.05 \\
    \end{tabular}
\end{table}

\subsection{Results and Discussion}

The time-domain plots of device frequency and real power output are presented in Fig. \ref{fig: Frequency for cases}, and Fig. \ref{fig: Power for cases}, respectively. Table \ref{tab: cases results} shows frequency statistics and power differentials from the steady state values for each case. 

\begin{table}[tb]
    \renewcommand{\arraystretch}{1.2}
    \setlength{\tabcolsep}{.3em}
    \centering
    \caption{Case Results--Frequency Statistics Apply to Synchronous Generator Shaft Speed; $\Delta P$ is machine per unit}
    \label{tab: cases results}
    \begin{tabular}{c||c|c||c|c}
        \textbf{Case} & \textbf{Nadir (Hz)} & \textbf{ROCOF (Hz/s)} & $\pmb{\Delta P_{SG}}$ & $\pmb{\Delta P_{GFM}}$\\
        \hline\hline
        A & 59.93 & 0.44 & 0.018 & 0.113 \\
        B & 59.82 & 0.61 & 0.041 & 0.069 \\
        C & 59.70 & 0.93 & 0.061 & 0.029 \\
    \end{tabular}
\end{table}

\begin{figure}[h]
    \centering
    \includegraphics[width=0.7\columnwidth,trim={0 0 0 38},clip]{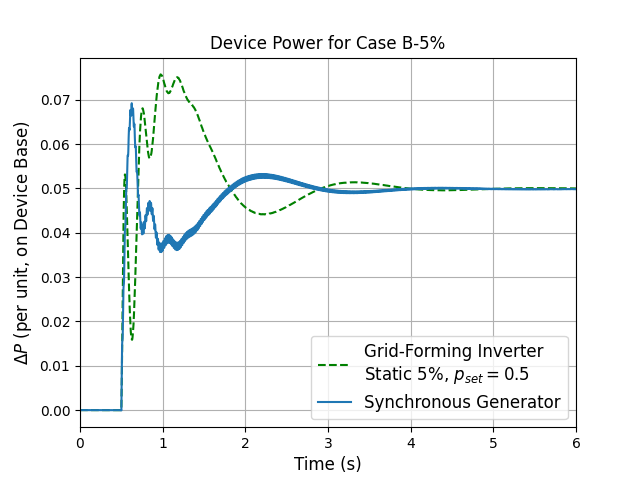}
    \caption{Synchronous generator and grid-forming inverter power output for static 5\% droop and case B dispatch. Response is identical for cases A, B, and C; A and C are not shown. Note that the grid-forming inverter power output exceeds 0.05 p.u., a power violation in case C.}
    \label{fig: 3bus power 5}
\end{figure}

The results of case A show slow frequency dynamics (Fig. \ref{fig: freq a}, with a ROCOF of 0.44 Hz/s and a nadir of 59.93 Hz, which is approximately equivalent to the settling frequency. The device powers (Fig. \ref{fig: power a}) show that the majority of the load perturbation was met by the GFM inverter. The slow frequency mode of $\lambda_{5,6}$ (explained in the previous section) is very well damped, as expected by the eigenvalue trajectory of Fig. \ref{fig: eigen 3}. The frequency response of both the SG and GFM are nearly identical. 

Case B, where the dispatch of the GFM inverter is $p_{set} = 0.5$, shows a relatively larger frequency deviation (Fig. \ref{fig: freq b}) as compared to case A, with the case B nadir 0.1 Hz lower. The ROCOF is larger, at 0.61 Hz/s, which is expected due to the larger \textit{Droop-e} gains as applied to \eqref{eq: GFM rocof}. A low frequency oscillatory element is more present in the response, which corroborates the decrease in damping at higher dispatches from the $\lambda_{5,6}$ trajectory. The power outputs of the devices show that a larger amount of power is delivered to the network from the SG, which is accomplished only because the frequency has a lower settling frequency. The GFM inverter still delivers a larger per unit quantity, indicative that even at a 50\% dispatch, the GFM can still contribute more power than for a static 5\% droop value, wherein the per unit power export of both devices would settle to the same value. 


Finally, the results of case C, when the \textit{Droop-e} GFM inverter is dispatched at $p_{m,I,set} = 0.95$, are shown in Fig. \ref{fig: freq c} and Fig. \ref{fig: power c}. First, the frequency dynamics show an even lower nadir (59.70 Hz), and a larger ROCOF of 0.93 Hz/s. Because the GFM inverter has very little headroom, its ability to diminish the frequency dynamics is minimal. The high frequency oscillatory mode of $\lambda_{1,2}$ is present in the GFM output; it is a local mode. These oscillations are exhibit poor damping; however, a mitigation strategy is beyond the scope of this paper and yield a direction for future research. Also present is the low frequency mode of $\lambda_{5,6}$ (explained in the previous section), which shows even less damping, further corroborating the eigenvalue trajectory of Fig. \ref{fig: eigen 3}. The power output of the devices includes a `GFM Limit' trace to show the maximum output permissible of the GFM device. Even though the device is dispatched very near the limit, at $p_{set} = 0.95$, the limit is not violated. This shows the naturally limiting behavior of the \textit{Droop-e} control, wherein the device maintains the frequency--power exponential relationship without additional PI controller action, as has been presented in other work \cite{lasseter_grid-forming_2020}. 

\FloatBarrier
\section{Power Sharing}
\label{sec: power sharing}

The \textit{Droop-e} control of the GFM is a strict departure from the static droop convention, which yields power sharing amongst frequency responsive devices. Namely, if all devices operating with frequency response maintain a global droop value (i.e., 5\% in North America), then all devices will contribute to power differentials equally, as a function of the device rating. If device do not share power equitably, then some may be far closer to limits than others, which can lead to instability. The \textit{Droop-e} control does not hold this power sharing objective, as the primary goal is to provide more power by maintaining smaller deviations in frequency via the nonlinear nature. Thus far, the efficacy of the \textit{Droop-e} control in mitigating frequency transients has been shown, but the question remains of how to achieve autonomous power sharing after the transients have diminished. Here, the unique capability of the GFM inverter to directly regulate frequency is used to develop a novel power sharing control.

\subsection{Power Sharing Controller}

\begin{figure}[h]
    \centering
    \includegraphics[width=0.8\columnwidth,trim={0 0 0 0},clip]{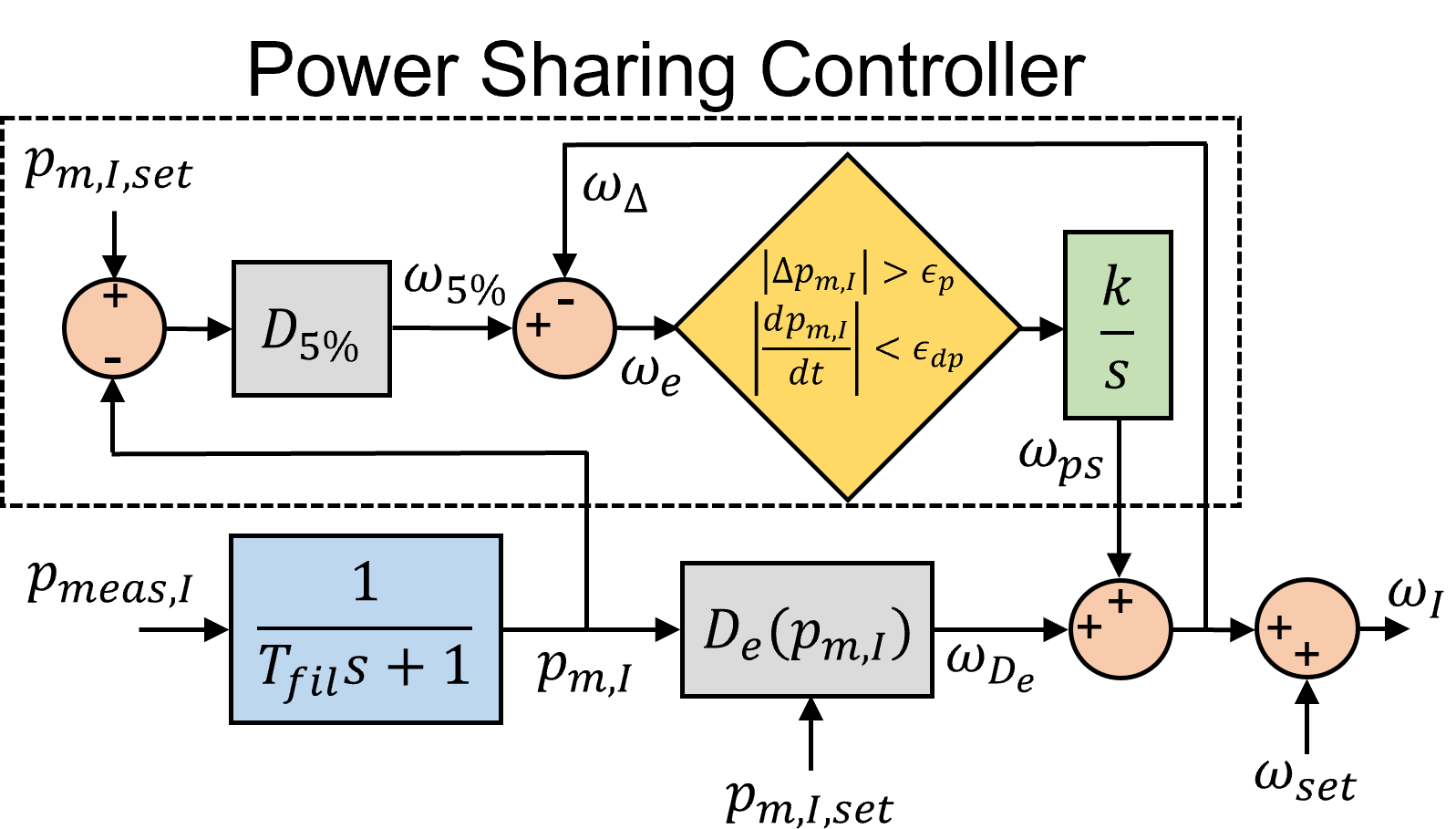}
    \caption{The power sharing controller that adds an offset, $\omega_{ps}$, to the output frequency $\omega_I$, to achieve autonomous, 5\% power sharing amongst other frequency responsive devices.}
    \label{fig: frequency recover}
\end{figure}

The proposed power sharing controller, presented in Fig. \ref{fig: frequency recover}, operates by modulating the output frequency with an offset component, $\omega_{ps}$. By not bypassing the fundamental \textit{Droop-e}, the GFM inverter will continue to provide damping to the system with the exponential droop relation, but will also change frequency such that the other frequency responsive devices react and equitable power sharing is accomplished. First, the frequency deviation that would result with a static droop (i.e., 5\%), $\omega_{5\%}$ in Fig. \ref{fig: frequency recover}, is directly calculated with the resultant power deviation \eqref{eq: static droop ps}. This frequency is compared with the \textit{Droop-e} output, $\omega_{D_e}$, and the resultant power sharing component $\omega_{ps}$, to generate an error \eqref{eq: freq error}.

\begin{align}\label{eq: static droop ps}
    \omega_{5\%} &= (p_{m,I,set}-p_{m,I})D_{5\%}\\
    \label{eq: freq error}
    \omega_{e} &= \omega_{5\%} - \omega_{\Delta} - \omega_{ps}
\end{align}


The logic block will remain open until a disturbance is registered \eqref{eq: disturbance criteria}:
\begin{equation}\label{eq: disturbance criteria}
    closed\phantom{0}if\phantom{0}
    \begin{cases}
        |\Delta p_{m,I}|>\epsilon_p \\
        |\frac{d p_{m,I}}{dt}| < \epsilon_{dp}
    \end{cases}
\end{equation}
where are $\epsilon_p$ and $\epsilon_{dp}$ are tolerance parameters. Once the disturbance criteria is met, this error is passed through an integrator block with gain $k$, which generates the frequency offset $\omega_{ps}$. As this offset is added to the output frequency $\omega_I$, $p_{m,I}$ will change due to the dynamics of AC power transfer and other frequency responsive devices on the network. This change is compensated for in the controller, and the GFM will arrive at the equitable, per unit power sharing value as $\omega_e$ is driven to $0$ by the integrator. 
Note that the static droop gain is a parameter that can be arbitrarily set; e.g., 4\% in Europe and 5\% in North America.


\subsection{Demonstration of Power Sharing Mechanism}

\begin{figure}[h]
    \centering
    \includegraphics[width=0.8\columnwidth,trim={0 4 40 38},clip]{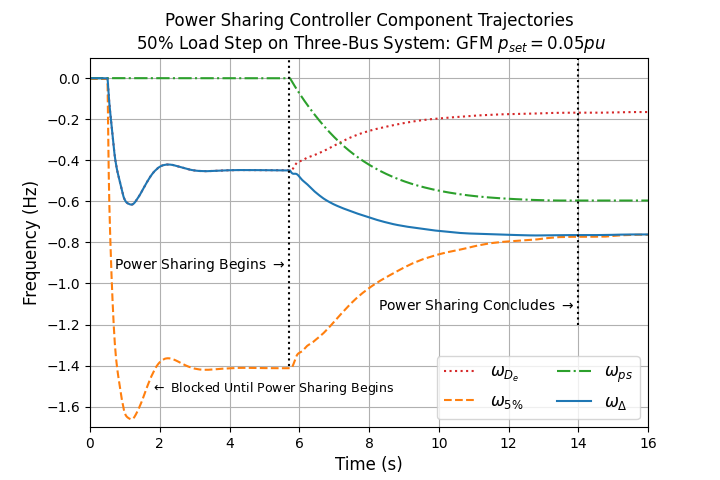}
    \caption{Power sharing controller variable trajectories during the power sharing interval of the simulation; $t=5.8-14s$. Variables match those from the controller diagram in Fig. \ref{fig: frequency recover}}
    \label{fig: frequency recover pieces}
\end{figure}

To demonstrate the efficacy of the power sharing controller in Fig. \ref{fig: frequency recover}, dynamic simulations were performed on the 3-bus system similar to the load steps of Section \ref{sec:simulation results}, but with a larger, 37.5 MW load increase (a 50\% increase, which is recognized as enormous, but used for illustrative purposes on this simple system) to show the capability of the \textit{Droop-e} control as well as that of the proposed power sharing strategy. Two simulations were performed; (i) with \textit{Droop-e} and (ii) with Static-5\% droop. The dispatch data corresponds to case A from Table \ref{tab: cases}. In the first simulation, one with \textit{Droop-e} control, $k = 0.3$. The results exhibit when the power deviation was registered ($|\Delta p_{m,I}| > \epsilon_p = 0.01\phantom{0}pu$), and the transients diminished ($|\frac{dp_{m,I}}{dt}<\epsilon_{dp} = \phantom{0}0.001$), the controller began applying the recovery offset, $\omega_{ps}$. Fig. \ref{fig: frequency recover pieces} displays the response contribution from different components involved in this power sharing control strategy, involving  $\omega_{D_e}$, $\omega_{5\%}$, $\omega_{\Delta}$, and $\omega_{ps}$. A factor of $(2\pi)^{-1}$ was applied to each trace for obtaining a Hz value. Once the logic gate was closed, the exponential change in $\omega_{ps}$ began. As the frequency of the device changed with $\omega_{ps}$, the output power $p_{m,I}$ also changed, which incurred changes in $\omega_{D_e}$ and $\omega_{5\%}$. At the conclusion of this extended controller action, the frequency successfully reached the equitable settling value with with $\omega_{\Delta}$ arriving at the $\omega_{5\%}$ value. 


\begin{figure}[h]	
	\centering
	\begin{subfigure}[t]{2.4in}
		\centering	\includegraphics[trim=0 4 40 38,clip,width=1\textwidth]{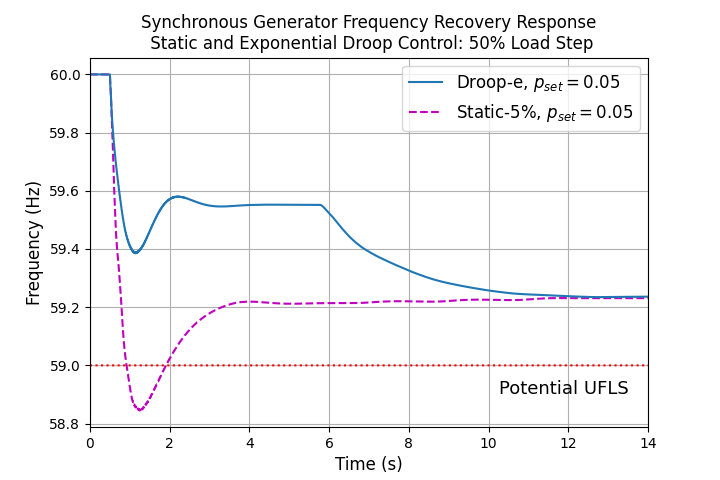}	\caption{Synchronous Generator Frequency Response}\label{fig: power recovery freq}		
	\end{subfigure}
	\quad
	\begin{subfigure}[t]{2.4in}
		\centering
		\includegraphics[trim=0 4 40 38,clip, width=1\textwidth]{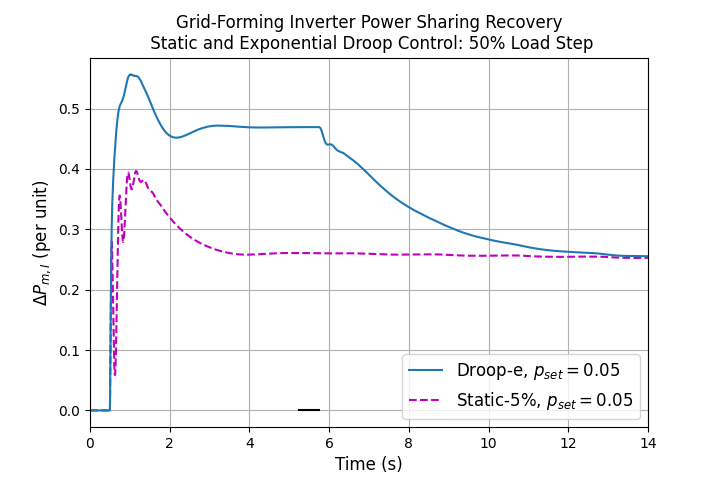}
		\caption{Grid-Forming Inverter Power Output}\label{fig: power recovery power}
	\end{subfigure}
	\caption{Power sharing recovery control implementation, following a 50\% load step. The static 5\% droop response is provided for comparison.}\label{fig: power sharing simulation}
\end{figure}

The novelty of the proposed power sharing controller allows the GFM device to compensate for power deficit and frequency variations in a more efficient way with \textit{Droop-e} control, and still settle at the very same value that a conventional 5\% static droop-based power sharing would yield. The results of these simulations are shown in Fig. \ref{fig: power sharing simulation}, with Fig. \ref{fig: power recovery freq} presenting the SG frequency response for these two simulations; the GFM frequency response with a static 5\% droop was nearly identical to the SG and hence, not shown. These frequency results corroborate the superiority of the power sharing extended \textit{Droop-e} control relative to the static droop control. They indicate the peak ROCOF for the \textit{Droop-e} control was 2.3 Hz/s, compared to 3.9 Hz/s for the static 5\% droop. The static 5\% case experienced a much more deviant frequency nadir, and entered potential UFLS territory; \textit{Droop-e} certainly did not. Once the power sharing controller was initiated, at approximately $t=6s$, the frequency response showed an exponential decrease (due to the integrator) as the GFM inverter tracked to achieve power sharing, and made GFM headroom available to respond to another potential event. The nadir with the \textit{Droop-e} control was the settling frequency.

Fig. \ref{fig: power recovery power} shows the power output of the GFM inverter for each controller; \textit{droop-e} vs. static 5\% droop. These results show that the \textit{Droop-e} control delivered more power to the network than the static 5\% droop control. When the power sharing controller was initiated, the power output exhibited a slow exponential decline to the 5\% droop value; equitable power sharing was achieved autonomously within 15 seconds of the perturbation, while this rate was a parameterized gain that can be tuned for a faster or slower response by adjusting the value of $k$.


\section{Time-Domain Simulations: IEEE 9-Bus System}
\label{sec: 9 bus}

\begin{figure}[h]
    \centering
    \includegraphics[width=0.75\columnwidth,trim={0 0 0 0},clip]{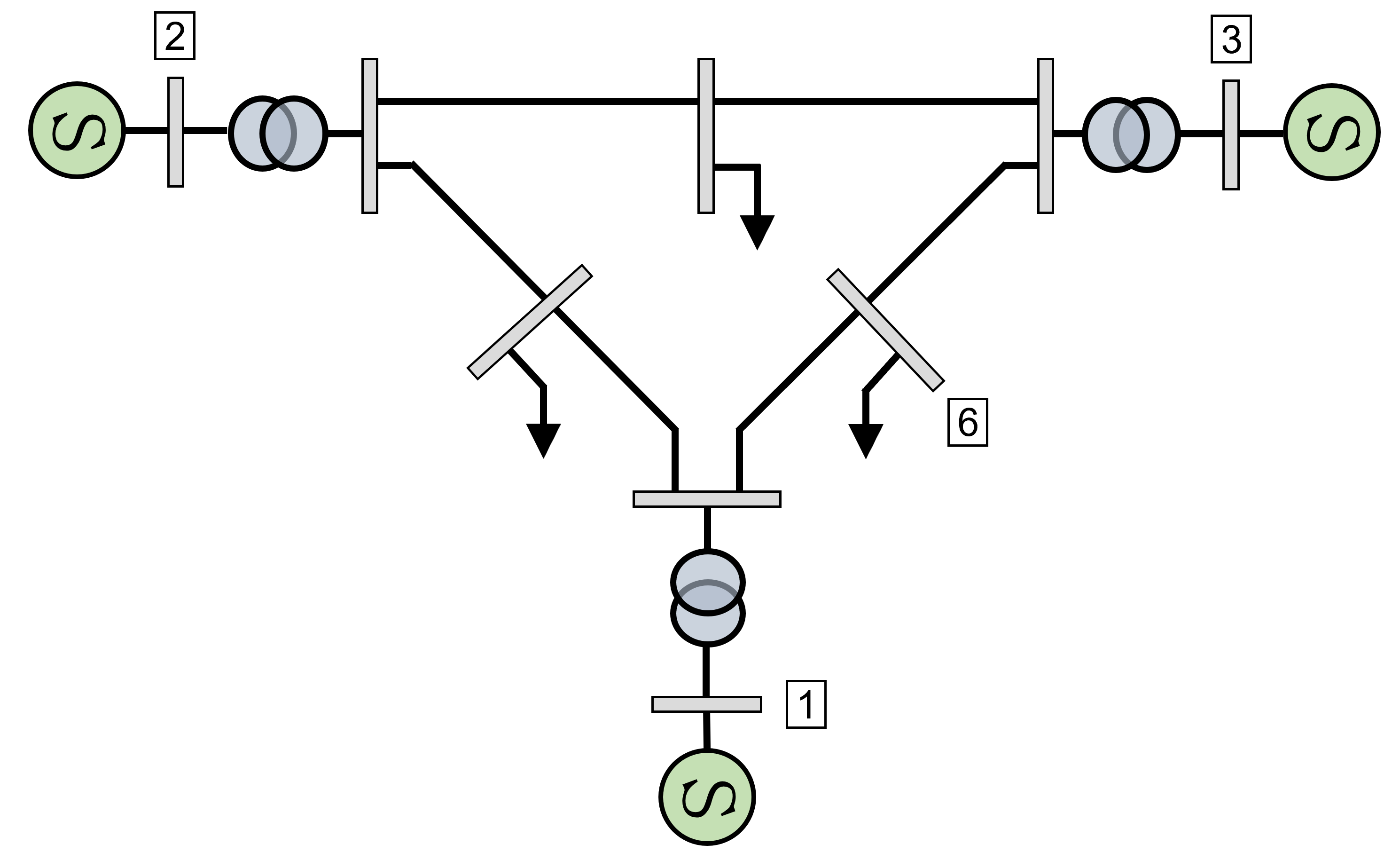}
    \caption{IEEE 9 bus system. Generators 1 and 3 are changed to Static-5\% and Droop-e in Cases 9-B, and 9-C, respectively.}
    \label{fig: 9 bus system}
\end{figure}

The IEEE 9-bus system \cite{manitoba_hydro_international_ltd_ieee_nodate-1}, as configured in \cite{kenyon_open-source_2021}, was used to demonstrate and validate the capability of the \textit{Droop-e} control on a mesh network, with multiple \textit{Droop-e} GFM devices. The system configuration is given in Table \ref{tab:9 Bus config}, which corresponds with the network diagram shown in Fig. \ref{fig: 9 bus system}. The perturbation applied was a 10\% load step at bus 6. Three cases were simulated, the first is 9-A with all generators as SGs, modelled as in the previous dynamic simulations. The second case, 9-B, has generators 1 and 3 supplanted with static 5\% droop GFMs. The third case, 9-C, has these two GFMs converted to \textit{Droop-e} control. The power sharing control parameters were the same as from Section \ref{sec: power sharing}.

\begin{table}[htbp]
    \small
    \centering
    \caption{9 Bus Configuration}
    \setlength\tabcolsep{3.5pt}   
    \begin{tabular}{c|c|c|c|c|c}
    \multirow{2}{*}{\rotatebox{0}{Generator}} & Rating & P & \multicolumn{3}{c}{Generator Type for Case}\\
     & (MVA) & (MW) & 9-A & 9-B & 9-C \\\hline\hline
    1 & 200 & 71.5 & SG & Static-5\% & Droop-e\\
    2 & 200 & 163 & SG & SG & SG\\
    3 & 200 & 85 & SG & Static-5\% & Droop-e\\
    \end{tabular}
    \label{tab:9 Bus config}
\end{table}

In these simulations, a weighted frequency is calculated according to $\label{eq:measurefrequency}
    f(t) = \frac{\sum_{i=1}^n (MVA_i*f_i(t))}{\sum_{i=1}^n MVA_i}
$ where $f_i(t)$ is the frequency of device $i$ at time $t$, $MVA_i$ is the device $i$ rating, and $n$ is the number of devices. This weighted frequency is used to determine the ROCOF and nadir values, according to the same definitions as presented in Section \ref{sec:simulation results}. The mechanical inertia rating of the system configuration, presented in Table \ref{tab: 9 cases results}, is a weighted average calculated as $\label{eq:aggregate inertia} H = \frac{\sum_{i=1}^n H_i S_{B,i}}{\sum_{i=1}^nS_{B,i}}$
where $H_i$ is the inertia rating (in $s$) of device $i$, $S_{B,i}$ is the MVA rating of device $i$, and $n$ is the number of devices. The inertia rating of the GFM devices is 0 s.

\begin{figure}[h]
    \centering
    \includegraphics[width=0.8\columnwidth,trim={8 8 35 37},clip]{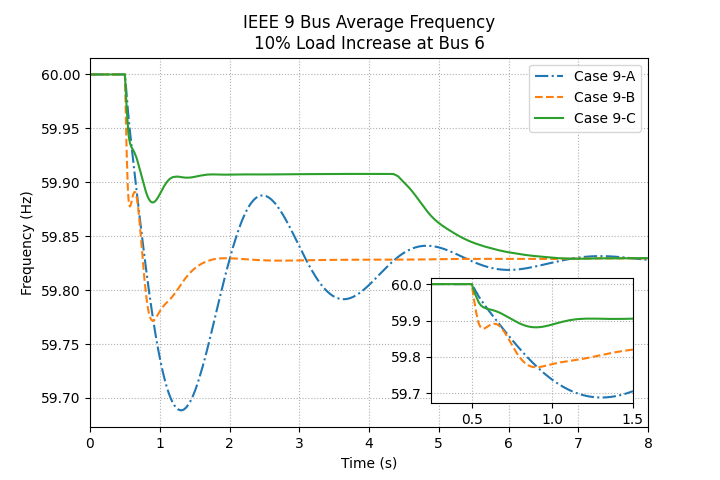}
    \caption{Average frequency response of 9-bus system for the three cases simulated following a 10\% load step at bus 6.}
    \label{fig: 9 bus freq}
\end{figure}

The time-domain average frequency response from the simulations of each case are presented in Fig. \ref{fig: 9 bus freq}. The blue, dot-dash, trace represents the Case 9-A response, with the quintessential second-order trajectory, the most deviant nadir (56.68 Hz) of the three cases, and an initial ROCOF of 0.69 Hz/s. Case 9-B, the orange dashed trace, shows an improvement in the nadir (59.77 Hz), but a much larger ROCOF of 1.22 Hz/s. Finally, Case 9-C, with the solid green trace, shows the superior performance of the \textit{Droop-e} control. In this case, Case 9-C, although the inertia is a third of Case 9-A, the ROCOF values are identical at 0.66 Hz/s. The frequency trace for 9-C showed only a small initial overshoot, but this did not even register as the nadir because of the relatively large immediate delivery of power due to the \textit{Droop-e} controller. It also showed approximately 4 seconds after the load step, the power sharing recovery control became engaged, with a gradual, exponential decrease in frequency towards to the settling frequency, 59.83 Hz, identical two the other two cases. The nadir for Case 9-C was the settling frequency. Table \ref{tab: 9 cases results} summarizes the frequency statistics for each case

\begin{table}[tb]
    \centering
    \caption{9-Bus Case Results--Frequency Statistics Derived From Average}
    \label{tab: 9 cases results}
    \begin{tabular}{c||c|c|c}
        Case & Nadir (Hz) & ROCOF (Hz/s) & Inertia (s)\\
        \hline\hline
        9-A & 56.68 & 0.69 & 3.0\\
        9-B & 59.77 & 1.22 & 1.0\\
        9-C & 59.83 & 0.68 & 1.0\\
    \end{tabular}
\end{table}

\begin{figure}[h]
    \centering
    \includegraphics[width=0.8\columnwidth,trim={8 8 5 34},clip]{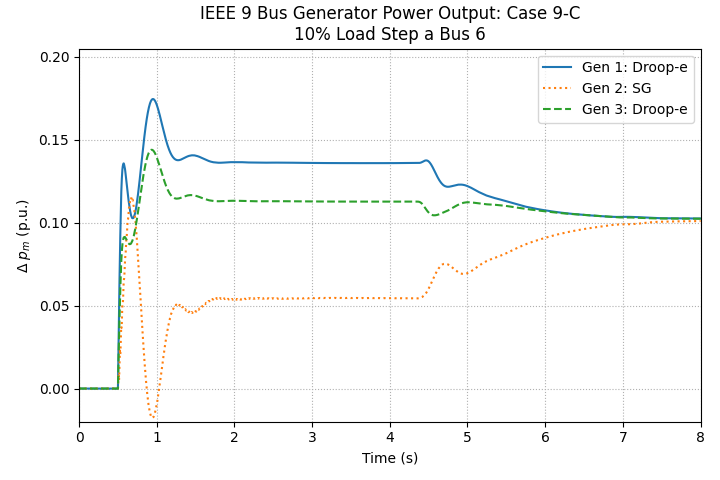}
    \caption{Individual device power responses for Case 9-C.}
    \label{fig: 9 bus power}
\end{figure}

The power response of each device for Case 9-C is shown in Fig. \ref{fig: 9 bus power}. It is clear that the two \textit{Droop-e} controlled GFM inverters at bus 1 and 3 delivered significantly more power to the network than the SG at bus 2. Consequently, the transients were diminished within 1.5 seconds of the load step perturbation. Due to the differences in dispatch and the operation of the \textit{Droop-e} control, the resultant power delivery from each GFM device was different. Approximately 4 seconds after the perturbation, the power sharing control was activated for each GFM, with Gen 3 engaging a tenth of a second prior to Gen 1. This control is autonomous and a factor of local variables; therefore, the time of initiation could vary amongst devices. The power output from all three devices then converged to an identical value, successfully achieving the 5\% droop derived contribution based on the size of the load step. 

The results here show a superior transient frequency response by \textit{Droop-e}, and the autonomous, equitable power sharing with multiple devices operating under the same control. 


\section{Conclusion}
\label{sec:conclusion}

This paper presented the novel \textit{Droop-e} control strategy for grid-forming inverters, which establishes an active power--frequency relationship based on an exponential function of the power dispatch. The advantages of this control approach consist of an increased utilization of available headroom, mitigated frequency dynamics, and a natural limiting behavior. The proposed controller was demonstrated and validated using both the small-signal stability analysis and computational time-domain EMT simulations and compared to the hitherto standard static droop approach. 
Further, a novel secondary control that achieves power sharing autonomously with multiple devices following the primary \textit{Droop-e} response to load perturbations was introduced and simulated on the 9-bus network. 
Some potential directions for future research are:
\begin{itemize}
    \item More comprehensive controller design to mitigate the high frequency mode present at high $p_{m,I,set}$ values.
    \item Stability analysis, both analytical and transient, of larger networks with multiple \textit{Droop-e} devices.
    \item Analysis of the secondary power sharing control with multiple devices \textit{Droop-e} devices.
    \item Investigation of \textit{Droop-e} on larger networks to explore the potential reduction in the quantity of frequency responsive devices required for standard contingencies.
\end{itemize}

{\footnotesize\section*{Acknowledgment}
This work was authored by the National Renewable Energy Laboratory, operated by Alliance for Sustainable Energy, LLC, for the U.S. Department of Energy (DOE) under Contract No. DE-AC36-08GO28308. 
The views expressed in the article do not necessarily represent the views of the DOE or the U.S. Government. The U.S. Government retains and the publisher, by accepting the article for publication, acknowledges that the U.S. Government retains a nonexclusive, paid-up, irrevocable, worldwide license to publish or reproduce the published form of this work, or allow others to do so, for U.S. Government purposes.}

\bibliographystyle{IEEEtran}
\balance
{\footnotesize\bibliography{Wallace_Lib}}
\end{document}